\documentstyle[aps,prbbib,twocolumn,epsf]{revtex}
\begin{document}
\draft
\title{Carbon nanotube-based quantum pump in the presence
of superconducting lead}
\author{Yadong Wei$^1$ and Jian Wang$^{2,a)}$}
\address{1. Department of Physics, School of Science,
Shenzhen University, Shenzhen 518060, China \\
2. Department of Physics, The University of Hong Kong,
Pokfulam Road, Hong Kong, China}
\maketitle
\begin{abstract}
Parametric electron pump through superconductor-carbon-nanotube based 
molecular devices was investigated. It is found that a dc current, which is 
assisted by resonant Andreev reflection, can be pumped out from such 
molecular device by a cyclic variation of two gate voltages near the 
nanotube. The pumped current can be either positive or negative under 
different system parameters. Due to the Andreev reflection, the pumped 
current has the double peak structure around the resonant point. The 
ratio of pumped current of N-SWNT-S system to that of N-SWNT-N system 
($I^{NS}/I^N$) is found to approach four in the weak pumping regime 
near the resonance when there is exactly one resonant level at Fermi 
energy inside the energy gap. Numerical results confirm that in the 
weak pumping regime the pumped current is proportional to the square 
of the pumping amplitude $V_p$, but in the strong pumping regime the 
pumped current has the linear relation with $V_p$. Our numerical results
also predict that pumped current can be obtained more easily by using
zigzag tube than by using armchair tube.
\end{abstract}


\newpage

\section{Introduction}

Physics of parametric electron pump has attracted great attention.
Classical pumps had been fabricated about a decade 
ago\cite{kouwenhoven,pothier}. Recently, Quantum-dot-based quantum pump 
has been the subject of both experimental\cite{switkes} and 
theoretical\cite{brouwer,aleiner,zhou,shutenko,wei1,aleiner1,avron1,levinson,brouwer2,buttiker1,wei2,renzoni,makhlin,chu,wang1,levinson1,entin1} 
investigations. The quantum pump generates current due to the cyclic 
variation of at least two system parameters while maintaining zero bias. 
As the charge pumped out of the system, it also produces the Joule heat 
along with the dissipation. Recently, Avron et al\cite{avron2} have 
derived the lower bound for the dissipation. This naturally leads to the
concept of optimal pump. To search for an optimal pump, the heat current
and the shot noise of quantum pump have been investigated using the 
time-dependent scattering matrix theory\cite{buttiker2,vavilov1,wbg2}.  
Very recently, to explain the experimentally observed 
anomaly\cite{switkes}, the finite frequency pumping theory has been 
developed\cite{vavilov,wbg1}. The adiabatic pumping theory has also been
extended to account for the Andreev reflection in the presence of 
superconducting lead\cite{wang1,blaauboer}, strong electron interaction 
in the Kondo regime\cite{wbg3}, and spin polarized pumped current 
when the ferromagnetic leads are present\cite{wu}. 
Due to the peculiar electronic properties of Carbon nanotube 
(CNT)\cite{yao,tsukagoshi,tans,frank,dai,cobden,hatem,chris,orli}, CNT-based
parametric electron pump has been investigated as a prototypical
nanometer-scale molecular device\cite{wei2}. It would be interesting to 
further explore how does Andreev reflection modify the quantum interference 
of CNT based quantum pump in the presence of superconducting lead. 
It is well known that in the presence of normal conductor-superconductor 
(NS) interface, an incoming electron-like excitation can be Andreev 
reflected as a hole-like excitation\cite{Andreev}. 
In this paper, we will study a hybrid structure where both carbon
nanotube and superconducting lead are present and examine the interplay
between the electronic properties of CNT and superconductivity.
Specifically, we investigate a parametric quantum pump that consists of 
a finite sized single wall carbon nanotube (SWNT) connected to one normal 
left lead (N) and one superconducting right lead (S), i.e., the parametric
quantum pump of N-SWNT-S system. Two pumping driving forces $X_1(t)$ and
$X_2(t)$ are established by applying cyclic, time-dependent voltages
to two metallic gates, which are capacitively coupled to the SWNT
(see inset of Fig.1). We have used nonequilibrium Green's function
approach\cite{jauho,sun1,wbg5} to calculate the pumped current. We found 
that in the presence of superconducting lead the pumped current is four
times as that of corresponding normal system in the weak pumping regime.
As the pumping amplitude increases, the dependence of pumped current 
crosses over from quadratic to the linear dependence. Due to the Andreev 
reflection, the pumped current exhibits double peak structure for single
resonant level in line with the chemical potential $\mu_s$ of the 
superconducting lead. When $\mu_s$ is in the middle of two resonant 
levels, another type of Andreev reflection occurs where an electron  
coming from normal lead tunnels via the lower resonant level and Andreev
reflected as a hole through the upper resonant level with a Cooper
pair created in the superconducting lead. For this two level Andreev 
reflection, the pumped current shows remarkable parity effect that
the direction of pumped current near one resonant level is opposite to
that of the other level. Finally, our numerical results show that
it is much easier to pump current through zigzag structure than armchair
structure.  

\section{Theory}

In this work, we assume that the variation of the pumping potentials are
very slow and the adiabatic approximation is appropriate. 
In this approximation, the pumped current flowing through 
the left normal metallic lead, in one cycle time $\tau$, is given 
by\cite{brouwer,wang1,foot1}
\begin{equation}
I^{NS}=\frac{q\omega}{2\pi} \int_0^\tau d\tau [\frac{dN_L}{dX_1}
\frac{dX_1}{dt} + \frac{dN_L}{dX_2} \frac{dX_2}{dt}]
\end{equation}
where the quantity $dN_L/dX_j$ is the partial density of states (DOS), 
called the injectivity\cite{GraButtiker,wang4}, of the left lead
\begin{eqnarray}
\frac{dN_L}{dX_j}=\frac{dN_L^e}{dX_j}-\frac{dN_L^h}{dX_j}
\end{eqnarray}
with
\begin{eqnarray}
\frac{dN_L^e}{dX_j}=-\int \frac{dE}{2\pi} (-\partial_E f)
Tr[G_{11}^r \Gamma_L G_{11}^{a} \Delta_j]
\end{eqnarray}
and
\begin{eqnarray}
\frac{dN_L^h}{dX_j}=-\int \frac{dE}{2\pi} (-\partial_E f)
Tr[G_{12}^r \Gamma_L G_{12}^{a} \Delta_j]
\end{eqnarray}
where $j=1$ or $2$. $dN^e_L/dX_j$ describes the number of electrons
coming from left lead and exiting the system as electrons due to the
external parameter $X_j$. $dN^h_L/dX_j$ describes the number of holes
coming from left lead and exiting the system as electrons due to the
external parameter $X_j$\cite{wang4}. $G_{11}$ and $G_{12}$ are the matrix
elements of the $2 \times 2$ Nambu\cite{Nambu} representation
and can be expressed as\cite{sun}:
\begin{eqnarray}
G_{11}^r(E)&=&\left[E-H_{d} - V_{pp} - {\bf \Sigma}^r_{11}
\right.\nonumber \\
&-&  \left.{\bf  \Sigma}_{12}^r  A^r  {\bf  \Sigma}_{21}^r
\right]^{-1}
\label{g11}
\end{eqnarray}
and
\begin{equation}
A^r= \left[E+H_{d}+V_{pp}-{\bf \Sigma}^r_{22} \right]^{-1}
\label{ar}
\end{equation}
Once the electron and hole Green's function $G^r_{11}$ and $A^r$ were
obtained, $G^r_{12}$ is calculated by
\begin{equation}
G_{12}^r = G_{11}^r {\bf \Sigma}^r_{12} A^r
\label{g12}
\end{equation}
Here $\Gamma_{L,R}=-2Im[\Sigma_{L,R}^r]$ is the line-width function and
${\bf \Sigma}^r={\bf \Sigma}^r_L + {\bf \Sigma}^r_R$ is the
total self energy given by 
\begin{equation}
{\bf \Sigma}^r_L= \left( \begin{array}{ll}
         \Sigma^r_L & ~~~ 0 \\
         0 & ~~~ -\Sigma^a_L
         \end{array}
  \right)
\end{equation}
where $\Sigma^r_\alpha \equiv P_\alpha - i \Gamma_\alpha/2$ is the self
energy of the lead $\alpha$ in the normal case. Here $P_\alpha$ is the 
real part and $\Gamma_\alpha$ is the linewidth function. 
The self energy for the superconducting lead is
\begin{equation}
{\bf \Sigma}^r_R=\left( \begin{array}{ll}
          P_R-i\Gamma_R\beta_1/2 & ~~~  i\Gamma_R\beta_2/2
\\
         i\Gamma_R\beta_2/2 & ~~~ -P_R-i\Gamma_R\beta_1/2
         \end{array}
  \right)
\end{equation}
where $\beta_1=\nu E/\sqrt{E^2-\Delta^2}$, $\beta_2=\nu \Delta/
\sqrt{E^2-\Delta^2}$ with $\nu=1$ for $E>-\Delta$ and $\nu=-1$ for 
$E<-\Delta$. Here $\Delta$ is the gap energy of superconducting lead and
chemical potential of superconducting lead $\mu_s$ has been set to zero. 
In the above equations, $H_d$ is the Hamiltonian of CNT. It is a $N 
\times N$ matrix, where $N$ is the total number of carbon atoms. $V_{pp}$ 
is a diagonal matrix describing the variation of the CNT potential 
landscape due to the external pumping potentials $X_1(t)$ and $X_2(t)$. 
In this work, we choose the two pumping potentials to be $X_1(t) \equiv 
V_1(t) =V_{10}+V_{1p}\sin(\omega t)$ and $X_2(t) \equiv V_2(t) =V_{20}+
V_{2p}\sin(\omega t+\phi)$, where $\phi$ is phase difference, $\omega$ 
is the pumping frequency and $V_{1p}$ and $V_{2p}$ are the pumping 
amplitudes. To simplify the numerical calculation, we mimic the gate 
effects by simply adding the potential $V_{pp}=V_1 \Delta_1 + V_2 
\Delta_2$ to the SWNT where $\Delta_i$ is the potential profile function 
and can be set to be unit for the gate region, zero otherwise. A more 
accurate study requires a numerical solution of the Poisson equation with
the gates providing the appropriate boundary conditions.

\section{Result and discussion}

We now apply Eq.(1) to calculate the current for the N-SWNT-S parametric 
pump. For simplicity, the SWNT is modeled with the nearest neighbor 
$\pi$-orbital tight-binding model with bond potential $V_{pp\pi}=-2.75$eV. 
This model gives a reasonable, qualitative description of the electronic 
and transport properties of carbon nanotubes\cite{ref1,marco}.
Recently, a S-SWNT-S device has been studied experimentally\cite{dai}. 
By tuning the transparency of the device, clear signals of Andreev 
reflection were observed. The dependence of the Andreev current on the 
device transparency, the behavior of the differential resistance in the 
sub-gap region, as well as the observed low-temperature resistance 
anomaly\cite{dai} can all be explained theoretically using the 
$\pi$-orbital tight-binding model\cite{wei4}. 
We assume that the SWNT is weakly coupled to the electrodes so that the 
pumping process is mediated by the resonant tunneling. We further assume
that strong electron-electron interactions may be neglected. 
Without losing generality, we set the energy gap of the 
superconducting lead to be $1.45meV$ (the gap of Niobium). We also 
apply the wide bandwidth limit for the self-energy\cite{jauho} and consider
the symmetric pumping, i.e.,  $V_{10}=V_{20}=V_0$ and
$V_{1p}=V_{2p}=V_p$. In the absence of pumping we have $V_p=0$ which
forms a symmetric double barrier with barrier height $V_0$
in the finite size nanotube. As a result, the discrete resonant levels 
are established within the double barrier structure. By adjusting $V_0$, 
we can control the positions of energy levels inside the energy gap
$\Delta$. Since the pumped current is proportional to $\omega$, we set 
$\omega=1$ for convenience. We also set $\hbar=2m=e=1$. When pumping 
frequency $\omega=100MHz$, which is close to the frequency used in 
Ref.\onlinecite{switkes}, the unit for the pumped current is 
$1.6 \times 10^{-11}$A. 
Finally, we do not consider the finite temperature effect and hence set 
temperature to zero. 

First, we consider an armchair (5,5) SWNT with 93 layers of carbon atoms
(total 930 atoms). The two metallic gates that provide the pumping
driving forces are located near the two ends of the SWNT from 10th to 
28th layer, and 66th to 84th layer. We have chosen $V_0 \approx 2.75V$ so 
that there is only one resonant level in the energy gap and the level 
is aligned with the chemical potential of the superconducting lead in
the absence of pumping voltage $V_p=0$. Fig.1 shows the transmission
coefficient versus Fermi energy $E_F$. For comparison, the transmission
coefficient for corresponding normal system (when $\Delta=0$) is also 
plotted. We see that for normal
system, the transmission coefficient (dashed line) has Lorentzian 
lineshape. In the presence of superconducting lead, the transmission 
peak (solid line) is flattened and narrowed. Fig.2 depicts the pumped 
current $I^{NS}$ versus the Fermi energy for different pumping amplitudes
$V_p$ with $\phi=\pi/2$ and $\Gamma_L=\Gamma_R=0.0136eV$. 
Several interesting observations are in order. (1). the pumped current
has large amplitude only near the resonant level showing clearly a
resonant assisting behavior. This is because the pumped current is
related to the global density of states (DOS) and the DOS of the  
system reaches its maximum near the resonant level\cite{wei1}.
(2). the amplitude of pumped current has double-peak structure 
around the resonant level. To understand this behavior, we plot the
Andreev reflection coefficient versus $E_F$ at different times $t$
during the pumping cycle in the right inset of Fig.2. The
Andreev reflection coefficient gives large value only around the
two pumping instants: $t=3\pi/4$ and $t=7\pi/4$ because at such
moments the energy level of the SWNT is just in line with the
chemical potential of the superconducting lead so that an excitation
of hole can be reflected when there is an incident election near the
Fermi surface, and vice verse. At other moments, Andreev reflection 
coefficient is very small and contributes little to the current integral 
in the time cycle. In the inset, we just plot Andreev coefficients at
certain moments: $t_0=3\pi/4$, $t_0=0.95 \times 3\pi/4$ (a little smaller 
than $t_0$) and $t_0=1.05 \times 3\pi/4$ (a little lager than $t_0$).  
We see that all three curves have double-peak feature (for the solid line
the double-peak is barely seen). To understand this, we examine the
Andreev reflection coefficient which takes on the Breit-Wigner form near
the resonant level $E_0$\cite{wei4}: 
\begin{equation}
T_A(E)=\frac{\Gamma_1^2\Gamma_2^2}{4\left[ 
E^2-E_o^2+\frac{\Gamma\delta\Gamma}{4} \right]^2 + \Gamma_1^2\Gamma_2^2
+E_o^2\left[ \Gamma+\delta\Gamma \right]^2 },
\label{Ta1}
\end{equation}
where $\delta\Gamma\equiv \Gamma_1-\Gamma_2$ and $\Gamma=\Gamma_1+
\Gamma_2$. Here $\Gamma_1$ and $\Gamma_2$ are linewidth functions which 
characterize the ability of electron tunneling through the left and 
right barriers, respectively. At the moment $t_0=3\pi/4$, the barrier 
heights for the left and right barriers are $V_0+(\sqrt{2}/2) V_p$ and 
$V_0-(\sqrt{2}/2) V_p$, respectively. In this case, we found that the 
position of the resonant level remains the same as when $V_p=0$, i.e., 
the resonant level is still at $E_0=0$. From Eq.(\ref{Ta1}), we have 
\begin{equation}
T_A(E)=\frac{\Gamma_1^2\Gamma_2^2}{4\left(
E^2+\frac{\Gamma\delta\Gamma}{4}\right)^2+\Gamma_1^2\Gamma_2^2 }
\ .
\label{Ta11}
\end{equation}
Eq.(\ref{Ta11}) indicates that if $\Gamma_1=\Gamma_2$ so that 
$\delta\Gamma=0$, then resonant Andreev reflection occurs at $E=0$ 
with $T_A(E=0)=1$. On the other hand, if $\Gamma_1 >\Gamma_2$ so that 
$\delta \Gamma > 0$, $T_A$ takes on a maximum value at $E=0$ but this 
maximum value is less than $1$. Furthermore, if $\Gamma_1 < \Gamma_2$ 
such that $\delta\Gamma <0$, $T_A$ is characterized by two resonant peaks 
with $T_A=1$ at energies $E_{\pm}=\pm \sqrt{-\Gamma\delta\Gamma}/2$. 
Since the left barrier is higher than the right one, we have
$\Gamma_1<\Gamma_2$. This explains why we have double-peak feature for
the solid curve in the right inset of Fig.2. 
At the moment $t_0=0.95 \times 3\pi/4$ or $t_0=1.05 \times 3\pi/4$, the
resonant level shift so that $E_0$ is no longer zero. The double-peak
structure can still be explained using Eq.(\ref{Ta1}).
Similar discussions apply to the moment $t=7\pi/4$.  
From the above discussion, it is clear that the double-peak feature of 
Andreev coefficient around $t_0=3\pi/4$ and $t_0=7\pi/4$ is responsible for
the behavior of the pumped current. (3). the pumped current
increases as the pumping amplitude $V_p$ increases (compare solid line,
dotted line, and dashed line in Fig.2).  (4). For comparison, we also plot 
the pumped current $I^N$ for N-SWNT-N system with $V_p=0.0014V$ 
(long dashed line in Fig.2).  We see that it gives only one broad peak 
with small magnitude rather than two peaks. This is because for normal 
system the transmission coefficient does not give two peaks at any pumping 
moment and the double-peak feature for transmission coefficient is the 
intrinsic feature solely due to the Andreev reflection at the NS interface. 
Due to the quantum interference between the direct reflection (with 
amplitude $E_d$) and the multiple Andreev reflection (amplitude $E_A$), 
the pumped current $I^{NS}$ of N-SWNT-S is greatly enhanced and is much 
larger than the pumped current $I^N$ of the same system when the 
superconducting lead becomes normal\cite{wang1}. Roughly speaking, we have
$I^{NS} \sim |E_d + E_A \exp(i\theta)|^2$ where $\theta$ is the phase 
difference between $E_d$ and $E_A$. In the right inset of Fig.2, we plot 
the ratio of $I^{NS}/I^N$ as a function of pumping amplitude $V_p$ at the 
resonant energy. At small pumping amplitude, the direct reflection and the 
multiple Andreev reflection are exactly in phase ($\theta=0$) and the 
complete constructive interference gives the ratio around four\cite{wang1}.  
As the pumping amplitude increases, the constructive interference effect 
in the N-SWNT-S system is suppressed because $\theta \neq 0$. 

Now we consider a zigzag (10,0) SWNT with $L=56$ layers of carbon atoms
(total 560 atoms). Two pumping driving forces are added on the
tube layers from 5th to 8th layer and from 49th to 52nd layer.
By adjusting $V_0 \approx 2.10065V$, one resonant level is available 
at $E_F=0$.  In the calculation, we set $\phi=\pi/2$ and $\Gamma_R=\Gamma_L
=0.0136eV$. The results are plotted in Fig.3. We see that the pumped 
current gives similar behavior to that of armchair SWNT system. The 
pumped current is large near the resonant level and also
has the double-peak structure around the resonant point.
The ratio of $I^{NS}/I^N$ is also about four in the weak pumping regime
(see the inset of Fig.3). Moreover there are several points worth
mentioning. First, the pumped current can either be positive or negative
even for the same SWNT but different energy levels or different
phase differences (see Fig.5 and Fig.6). Comparing with armchair structure, 
the pumped current for zigzag structure reverses the direction. This is
because the pumped current is not due to the external bias but cause
by the pumping potentials. As a result, the pumped current is very sensitive
to the system parameters. Second, the double-peak for the pumped current 
is asymmetric especially when the pumping amplitude is large (see the 
dashed line of Fig.3). This is mainly due to the energy dependence of the 
self-energy. Third, under the same system parameters\cite{foot2}, the 
pumped current of zigzag structure is much larger (at least ten times 
larger) than that of armchair structure. This means that in order to 
obtained large pumped current, one should use zigzag nanotube instead of 
armchair nanotube. This may be useful for experimental study of the carbon 
nanotube pump. 

Fig.4 gives the pumped current as a function of pumping amplitude $V_p$ 
at $E_F=0$ for armchair SWNT. Here the system parameters are the same as 
those in Fig.2. We see that the pumped current increases quadratically at
small pumping amplitude and then reaches linear regime for large pumping 
amplitude. In order to understand this figure, we also plot $I=-1.06 
\times 10^4 V^2_p$ in the same plot. We confirm that in the weak pumping 
regime the pumped current is proportional to the square of the pumping 
amplitude, but in the strong pumping regime the pumped current is 
linearly proportional to $V_p$ only. Similar conclusion can be drawn from
the zigzag nanotube (see the inset of Fig.4). 
Fig.5 and the inset of Fig.5 present the pumped currents of armchair SWNT
versus phase difference at different pumping amplitudes in the weak pumping
regime and the strong pumping regime, respectively. We see that 
the pumped current is anti-symmetric about $\phi=\pi$, just like the
the result given by the experiments of Ref.\cite{switkes}
although superconducting lead was not used there. 
In the weak pumping regime, the sinusoidal behavior is clearly seen. 
In the strong pumping regime, however,  we see strong deviation from the
sinusoidal behavior (see inset of Fig.5) and the maximum pumped current
occurs near $\phi=\pi$ instead of $\phi/2$. 
 
Finally, we examine another N-SWNT-S quantum pump using a zigzag (10,0)
tube with $L=92$ layers (total atoms 920). One gate is located from 10th
to 28th layer and the other is located from 65th to 83rd layer. By adjusting
$V_0 \approx 2.5936V$, we obtain two double degenerated resonant levels at
$E_1=-3.5495 \times 10^{-5}eV$ and $E_2=3.5495 \times 10^{-5}eV$. Hence,
large Andreev reflections can occur near $E_F=E_1$ and $E_F=E_2$ with 
transmission coefficient equals to two (see the upper panel of Fig.6). 
Here the Andreev reflection is due to different origin from that of
Fig.1. In Fig.1, we have $E_0=0$ and Eq.(\ref{Ta1}) gives us the
resonance at $E=0$ with $T_A=1$. If $E_0$ is nonzero the maximum Andreev
reflection is less than one. In the strong tunneling regime which
applies to our case, $\Gamma_1$ and $\Gamma_2$ are very small. To
simplify the discussion, let us assume $\Gamma_1=\Gamma_2 << 1$, then from 
Eq.(\ref{Ta1}), we have
$$T_A = \frac{\Gamma_1^4}{[4(E^2-E_0^2)^2+\Gamma_1^4+4 E_0^2 
\Gamma_1^2]}$$
At resonance, $T_A = \Gamma_1^2/[\Gamma_1^2+4 E_0^2]$. Hence when $E_0$
is larger than $\Gamma_1$, the Andreev reflection quickly goes to zero.
However, if the chemical potential of the superconducting lead ($\mu_s$=0
in our case) is right in the middle of two resonant levels 
($E_1$ and $E_2$), {\it i.e.}, $\mu_s = (E_1+E_2)/2$, then electron coming from 
normal lead with incident energy $E_1$ tunnels into the structure through the 
resonant level $E_1$ and Andreev reflected as a hole back to the quantum 
dot through the resonant level $E_2$ with a Copper pair created in the 
superconducting lead, giving rise to the complete transmission. This is 
why in the upper panel of Fig.6 we have two transmission peaks with
$T_A=2$ due to the double degeneracy. 
The pumped current as a function of Fermi energy is plotted in Fig.6.
For comparison, we also plot the transmission coefficient (long
dashed line for normal structure and solid line for NS structure) in the 
upper panel of Fig.6. The pumped current with $V_p=1 \times 10^{-6} V$ 
(long dashed line in Fig.6) for the N-SWNT-N system with the same system 
parameters is also shown. Similar to Fig.2 and Fig.3, the pumped current 
also shows strong resonant behavior. It has large value near energies where 
the Andreev reflection peaks occur, while it diminishes quickly away from 
the peaks. The amplitude of pumped current also increases as the pumping 
amplitude $V_p$ increases. 
The striking feature here is that the pumped current peaks have
opposite sign for the two energy levels. That means the pump has
property that the DC current can flow out of the device from either
electrodes by a slight change of electron energy. 
From the lower inset of Fig.6, we see that the pumped
current clearly consists of two asymmetric peaks. 
Finally, we notice that large pumped current is generated for very small
pumping amplitude (compare Fig.3 with Fig.6). This is because in Fig.6
the thickness of potential barrier and hence the effective potential 
barrier height is much higher than that in Fig.3. As shown in
Ref.\onlinecite{levinson,wang2} that the maximum pumped current can reach
$1/2\pi$ for extremely large barrier and in strong pumping regime. 

In summary, we have investigated the parametric pump of N-SWNT-S
systems. By comparing with the parametric pump of N-SWNT-N system,
we find that in the presence of superconducting lead, the pumped
currents is greatly enchanced due to the quantum interference of
direct reflection and multiple Andreev reflection. In the weak
pumping regime, the pumped current is proportional to
the square of the pumping amplitudes but in the strong pumping regime,
the dependence becomes linear. Hence large pumped current can be
generated by increasing the pumping amplitude. When two level Andreev
reflection occurs, the pumped current show remarkable parity effect
so that the pumped current at one resonant level has opposite direction
of that of the other resonant level. Our numerical results 
also show that the zigzag nanotube is a better candidate for the pumping 
device since it is more easily pumped than the armchair nanotube.
In view of the S-SWNT-S structure studied in Ref.\onlinecite{dai}, it is
conceivable that the N-SWNT-S quantum molecular pump can be realized 
experimentally and the results presented here be verified.

\section*{Acknowledgments}
We gratefully acknowledge support by a RGC grant from the SAR Government of 
Hong Kong under grant number HKU 7091/01P.

\bigskip

\noindent{$^{a)}$ Electronic mail: jianwang@hkusub.hku.hk}

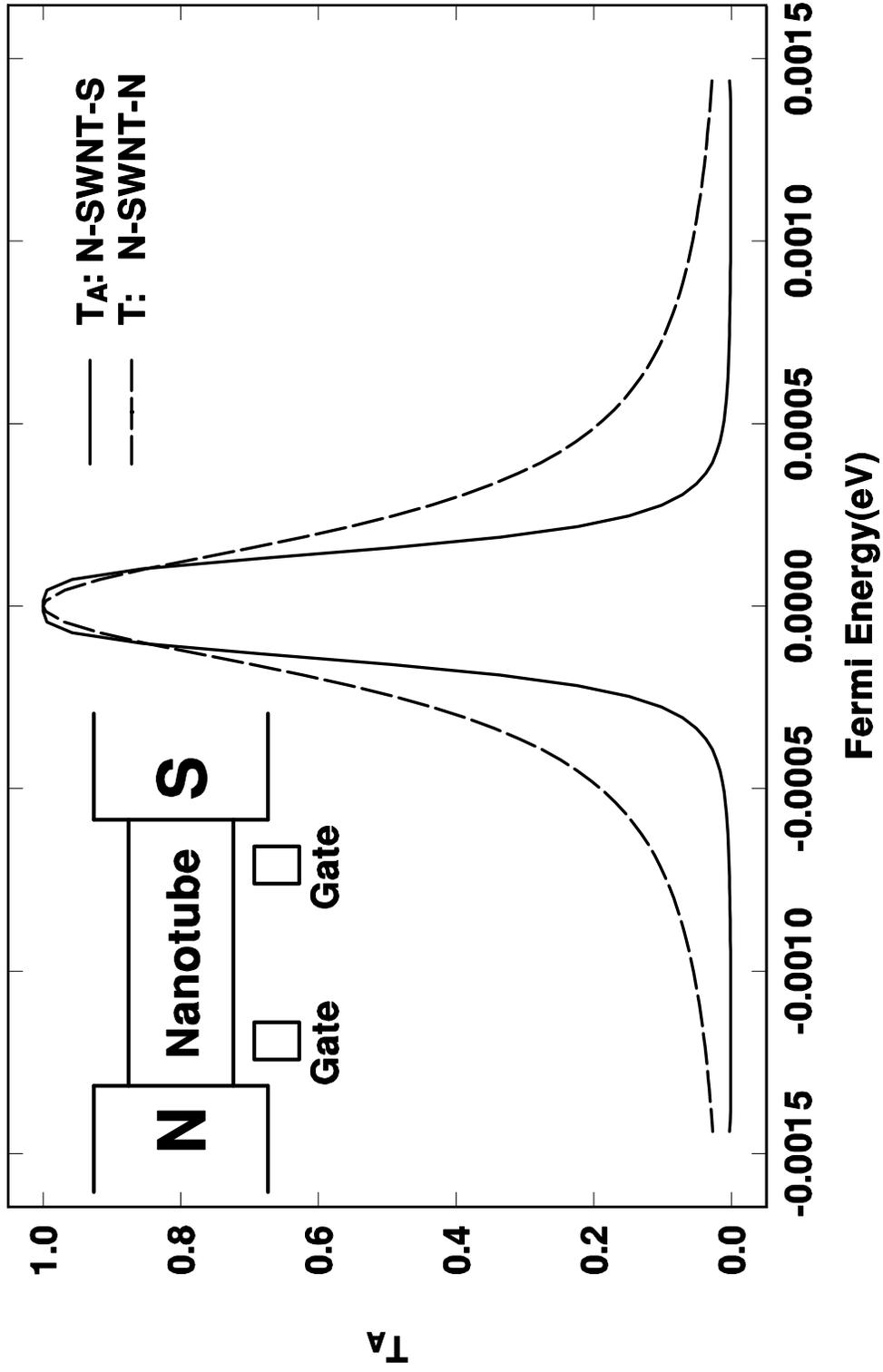
\begin{figure}[th]
\caption{
Andreev reflection coefficient $T_A$ as a function of
Fermi energy for N-SWNT-S system (solid line) and transmission
coefficient $T$ as a function of Fermi energy for
N-SWNT-N device (long dashed line). The SWNT is an armchair (5,5)
metallic tube.  
Inset: a schematic plot of the molecular device.
}
\end{figure}

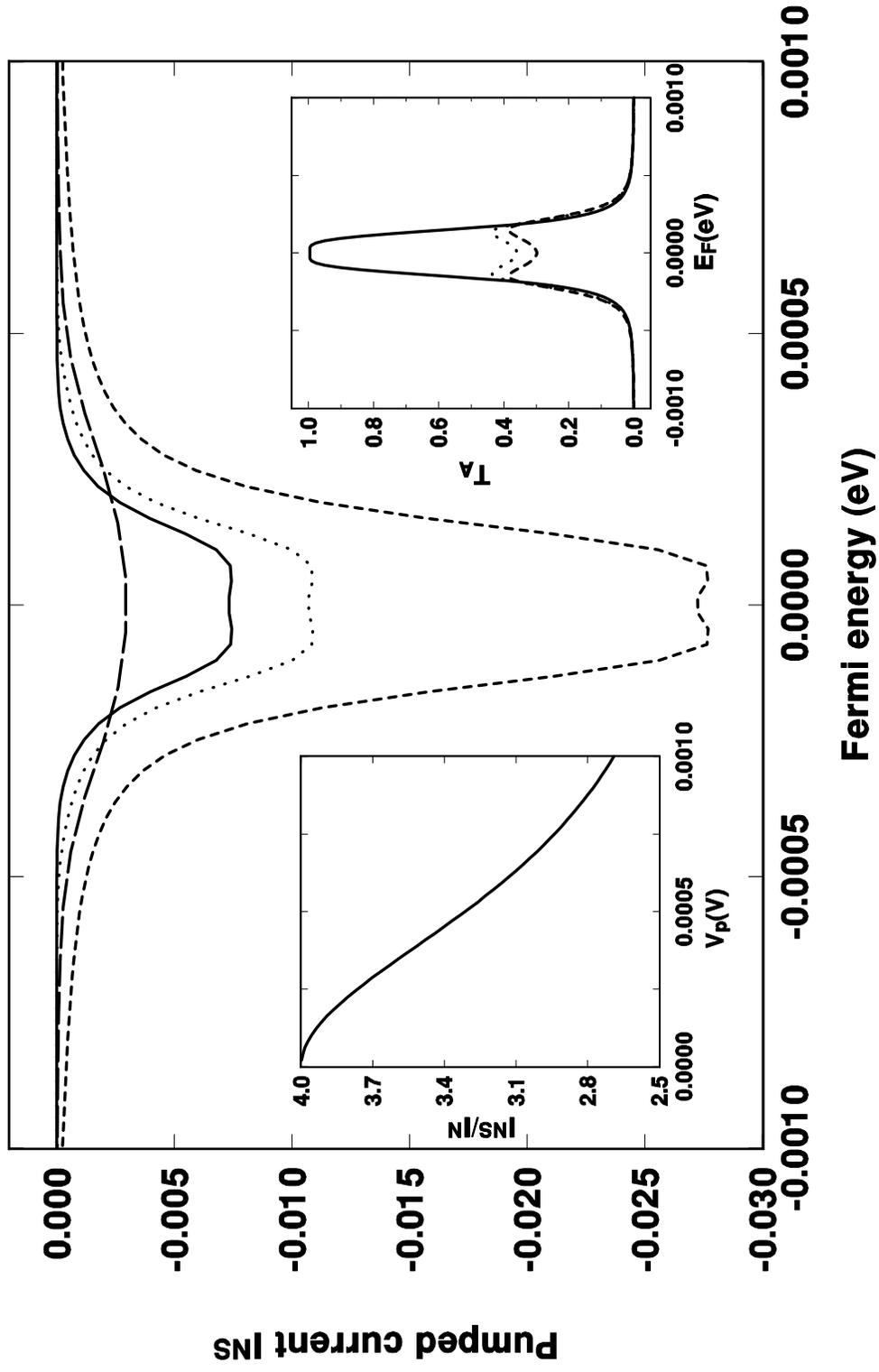
\begin{figure}
\caption{
The pumped current $I^{NS}$ versus Fermi energy for N-SWNT-S device at
different pumping amplitudes: $V_p=0.0014V$ (solid line), $V_p=0.002V$
(dotted line) and $V_p=0.005V$ (dashed line). Here long dashed line is 
the pumped current $I^N$ versus Fermi Energy for the corresponding 
N-SWNT-N device (when $\Delta=0$) at $V_p=0.0014V$. The left inset: 
$I^{NS}/I^N$ versus $V_p$ at resonant point. The right inset: Andreev 
reflection coefficient $T_A$ versus $E_F$ with $V_p=0.005V$ at different 
pumping moments: $t=3\pi/4$ (solid line), $t=0.95 \times 3\pi/4$ (dotted 
line) and $t=1.05 \times 3\pi/4$ (dashed line). 
}
\end{figure}

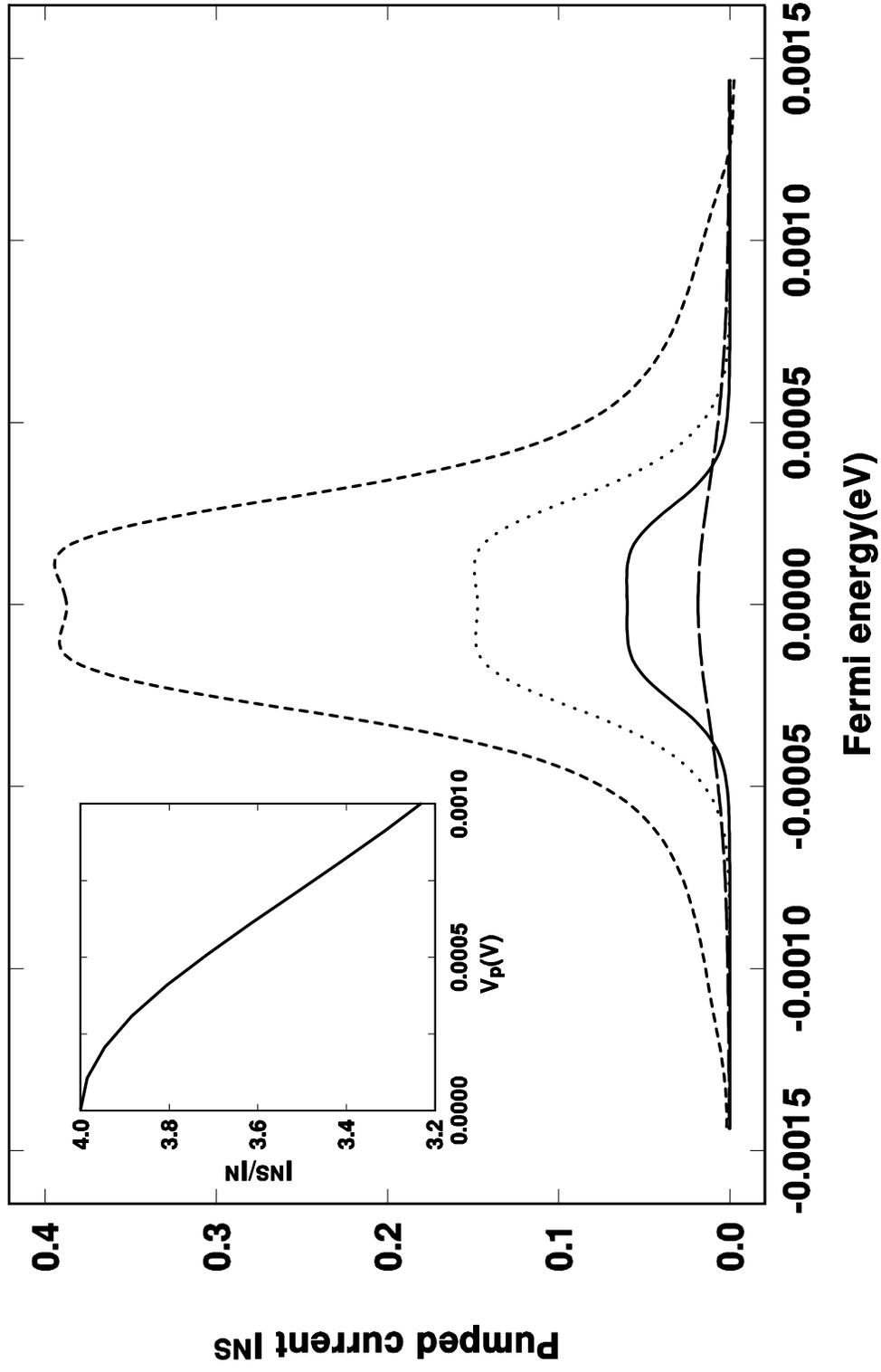
\begin{figure}
\caption{
The pumped current $I^{NS}$ versus Fermi Energy for N-SWNT-S device at
different pumping amplitudes: $V_p=0.001V$ (solid line), $V_p=0.002V$
(dotted line) and $V_p=0.005V$ (dashed line). The long dashed line is 
for the pumped current $I^N$ versus Fermi Energy of the corresponding 
N-SWNT-N device at $V_p=0.001V$. Here the SWNT is a zigzag (10,0) 
nanotube. The left inset: $I^{NS}/I^N$ versus $V_p$ at resonant point.
}
\end{figure}

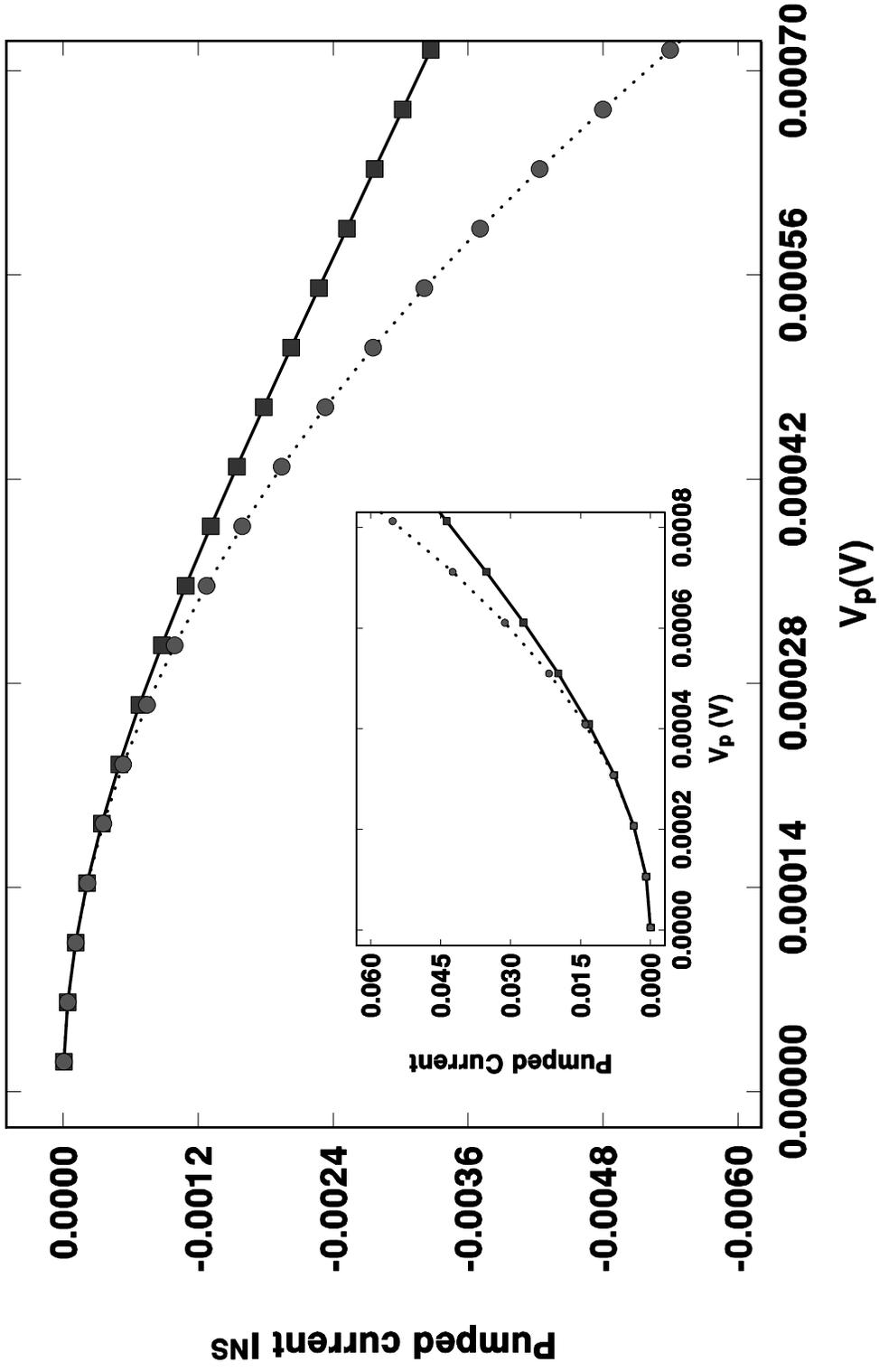
\begin{figure}
\caption{
The pumped current $I^{NS}$ of the armchair SWNT as a function of pumping 
amplitude $V_p$ at $E_F=0$ (solid line). Other system parameters are the 
same as those in Fig.2. Dotted line is plotted according to $I=-1.06 
\times 10^4 V^2_p$. Inset: The pumped current $I^{NS}$ of the zigzag
SWNT as a function of pumping amplitude $V_p$ at $E_F=0$ (solid line).
System parameters are the same as those in Fig.3.
Dotted line is plotted according to $I=8.3758 \times  10^5 V^2_p$.
}
\end{figure}

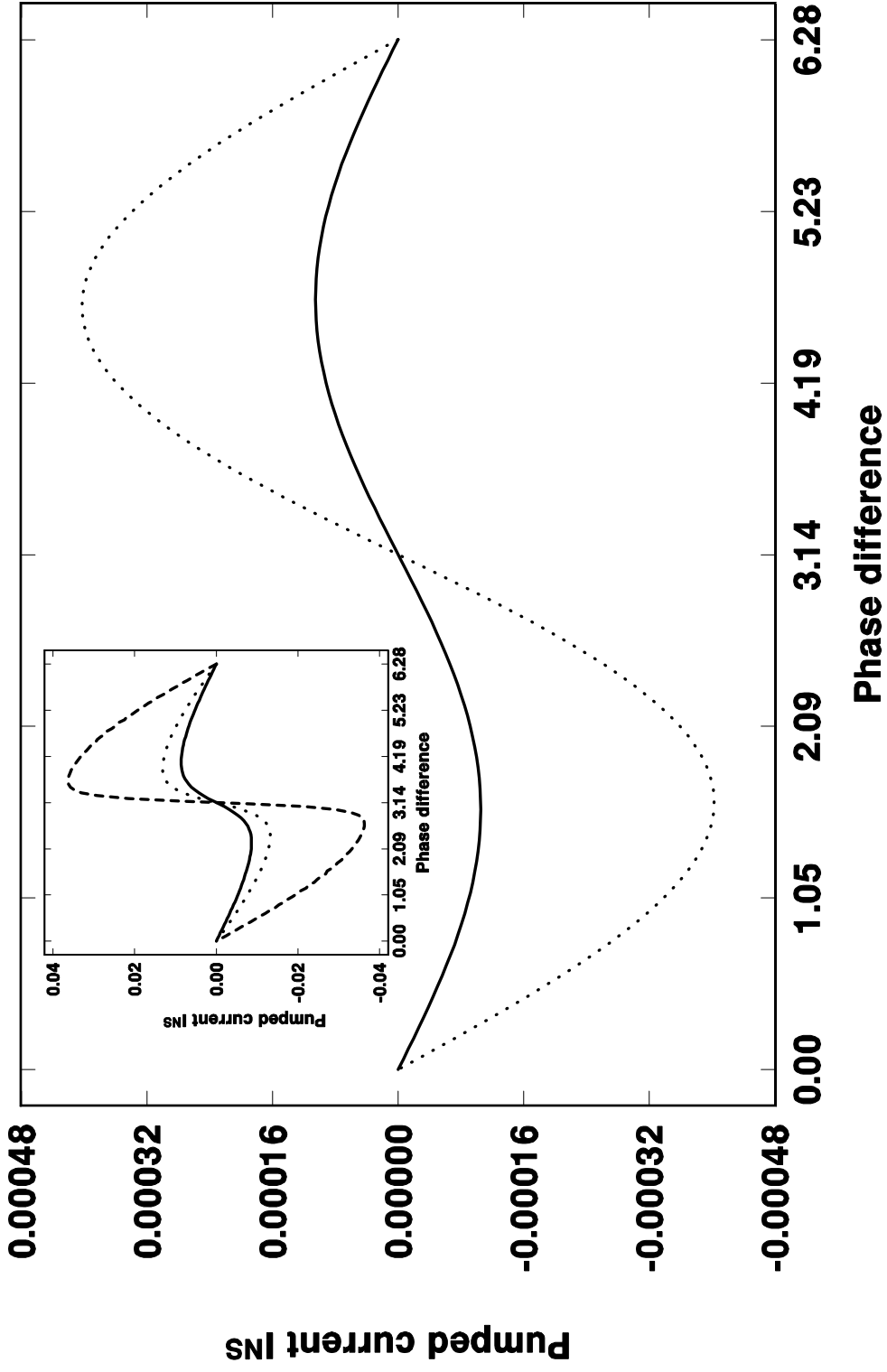
\begin{figure}
\caption{
The pumped current $I^{NS}$ of the armchair SWNT as a function of phase 
difference $\phi$ at $E_F=0$. Main figure: the weak pumping regime
with $V_p=0.0001V$ (solid line) and $V_p=0.0002V$ (dotted line). 
Inset: the strong pumping regime with $V_p=0.0014V$ (solid line), 
$V_p=0.002V$ (dotted line), and $V_p=0.005V$ (dashed line). 
Other system parameters are the same as those in Fig.2.
}
\end{figure}

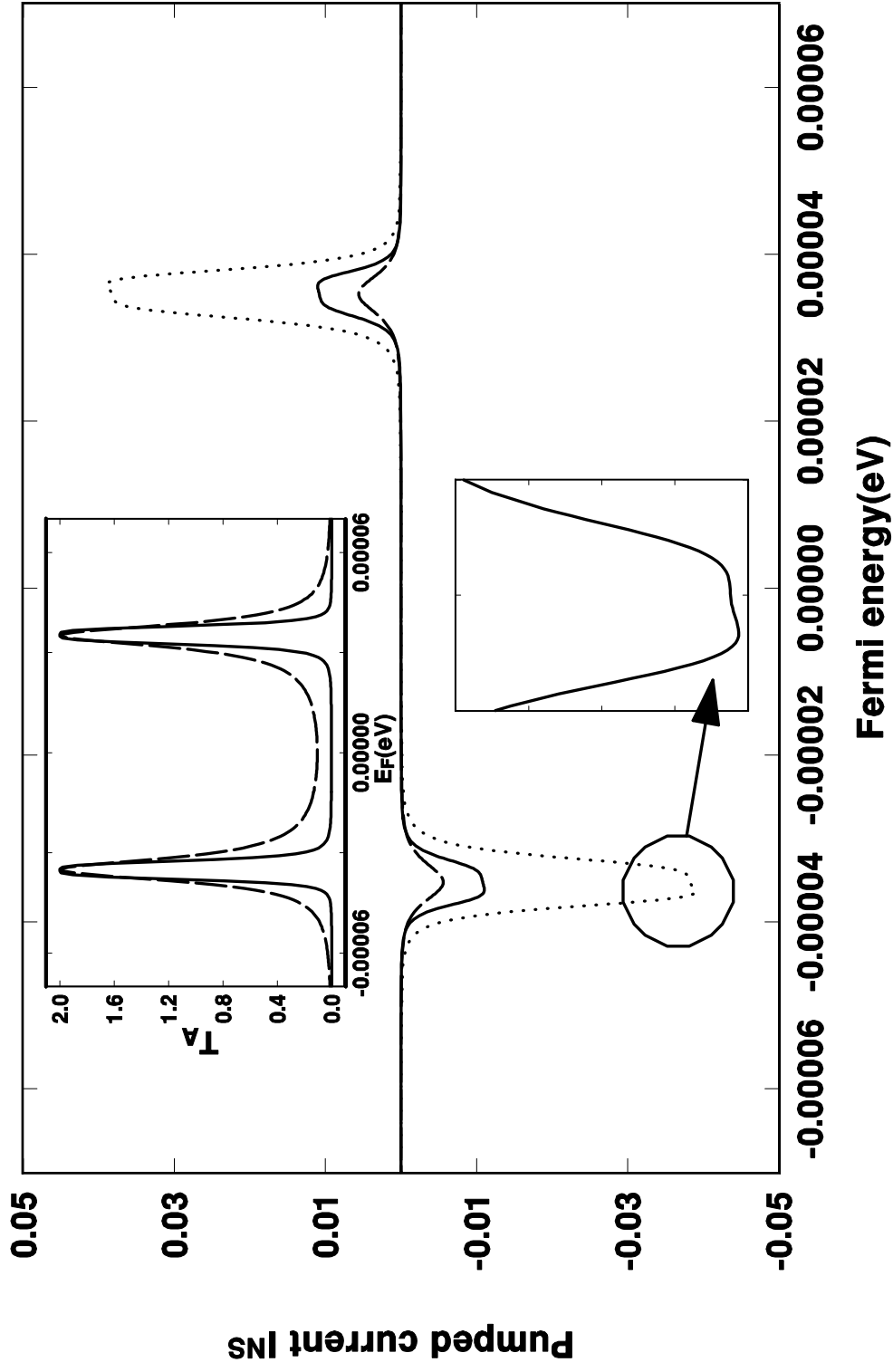
\begin{figure}
\caption{
The pumped current $I^{NS}$ versus Fermi Energy for the N-SWNT-S device at 
different pumping amplitudes: $V_p=1 \times 10^{-6}V$ (solid line),
$V_p=2 \times 10^{-6}V$ (dotted line). The long dashed line is the pumped 
current $I^N$ versus Fermi Energy for the corresponding N-SWNT-N device 
at $V_p=1 \times  10^{-6}V$. The SWNT is a zigzag (10,0) nanotube with
length L=92 layers.  The upper inset: Andreev reflection coefficient $T_A$ 
for N-SWNT-S device (solid line) and transmission coefficient $T$ for
the corresponding N-SWNT-N device (long dashed line). The lower inset:
the amplified figure of a pumped current peak.  The other system 
parameters: $\phi=\pi/2$ and $\Gamma_R=\Gamma_L=0.0136eV$.
}
\end{figure}


\begin{thebibliography}{00}
\bibitem{foot1}
We have omitted the factor of two for spin.
\bibitem{foot2}
In fact, the barrier thickness and barrier height of armchair nanotube
are larger than that of zigzag nanotube which favor the pumping of 
armchair nanotube. Yet the pumped current for zigzag nanotube is much 
larger than that of armchair nanotube with the same pumping amplitude.
\end{thebibliography}
\end{document}